# Exploring OpenStreetMap Availability for Driving Environment Understanding

Yang Zheng, Izzat H. Izzat, and John H.L. Hansen, *Fellow, IEEE*

*Abstract* — **With the great achievement of artificial intelligence, vehicle technologies have advanced significantly from human centric driving towards fully automated driving. An intelligent vehicle should be able to understand the driver's perception of the environment as well as controlling behavior of the vehicle. Since high digital map information has been available to provide rich environmental context about static roads, buildings and traffic infrastructures, it would be worthwhile to explore map data capability for driving task understanding. Alternative to commercial used maps, the OpenStreetMap (OSM) data is a free open dataset, which makes it unique for the exploration research. This study is focused on two tasks that leverage OSM for driving environment understanding. First, driving scenario attributes are retrieved from OSM elements, which are combined with vehicle dynamic signals for the driving event recognition. Utilizing steering angle changes and based on a Bi-directional Recurrent Neural Network (Bi-RNN), a driving sequence is segmented and classified as lane-keeping, lane-change-left, lane-change-right, turn-left, and turn-right events. Second, for autonomous driving perception, OSM data can be used to render virtual street views, represented as prior knowledge to fuse with vision/laser systems for road semantic segmentation. Five different types of road masks are generated from OSM, images, and Lidar points, and fused to characterize the drivable space at the driver's perspective. An alternative data-driven approach is based on a Fully Convolutional Neural Network (FCNN), OSM availability for deep learning methods are discussed to reveal potential usage on compensating street view images and automatic road semantic annotation.**

## I. Introduction

DRIVING is a comprehensive task that incorporates information among the driver, vehicle, and environment. Drivers make judgement based on the perception of environment, and then perform executions to control the vehicle. With recent advancements in artificial intelligence, vehicle technologies have progressed significantly from the human driving towards fully automated driving. During this transition, the intelligent vehicle should be able to understand the driver's perception of the environment and controlling behavior of the vehicle, as well as provide human-liked interaction with the driver. Next generation vehicles will need to be more active in assessing driver awareness, vehicle capabilities, and traffic and environmental settings, plus how these factors come together to determine a collaborative, safe and effective driver–vehicle engagement [1].

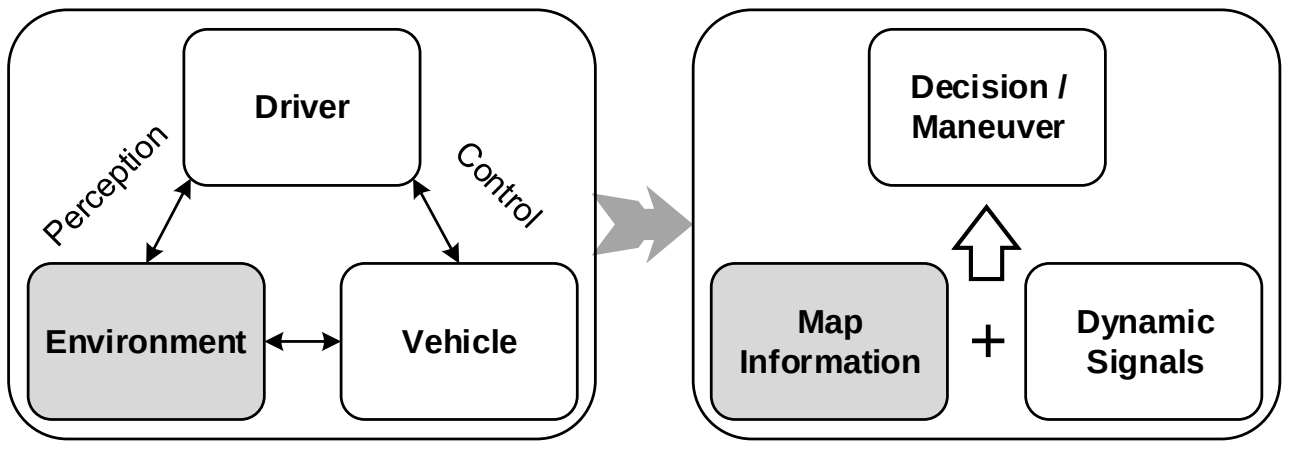

Fig. 1. Use map data to retrieve environment information, and combine with vehicle dynamic signals to understand driver behavior.

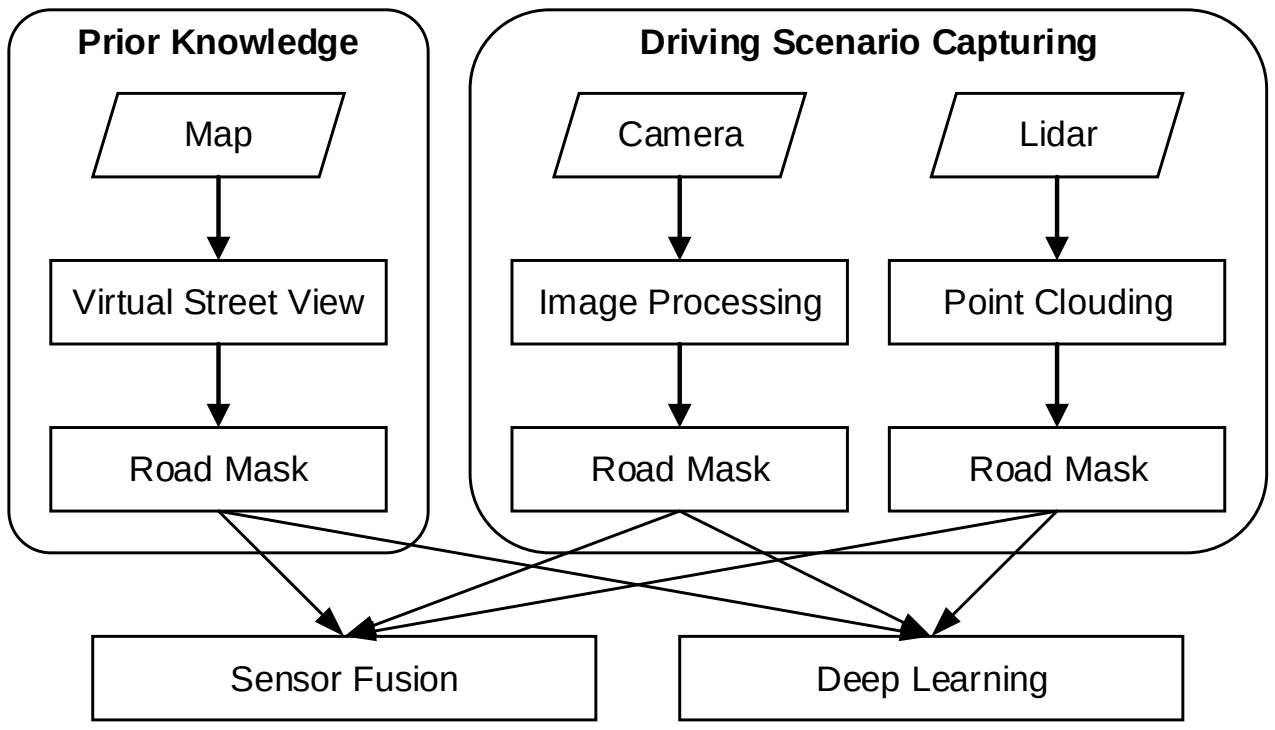

Fig. 2. Use map data to obtain prior road mask, and fuse with image and Lidar points for road semantic segmentation.

Today, high digital map databases (e.g. Google Map, HERE Map, OpenStreetMap, etc.) are available to provide rich environmental information such as static roads, buildings and traffic infrastructures. Driving scenario attributes can be extracted from such map data, including speed limit, number of lanes, distance to intersection, road curvatures, and so on. This traffic/environmental information is useful for the vehicle to understand how the driver makes decisions, and therefore contributing to the automated vehicle control and motion planning. To understand driving activity, one typical approach is to identify the driving context at micro-level events, and then assess their variations against expected patterns [2, 3]. However, it is challenging to characterize driving events from

Yang Zheng and John H.L. Hansen are with the Center for Robust Speech System (CRSS) – UTDrive Lab in the Electrical Engineering Department, University of Texas at Dallas, Richardson, TX 75080 USA (e-mail: {yxz131331, john.hansen}@utdallas.edu).

Izzat H. Izzat is with the Advanced Perception Group of Aptiv Inc., Agoura Hills, CA 91301 USA (e-mail: Izzat.izzat@aptiv.com)

TABLE I
MAP DATA COMPARISON

| | ***Google*** | ***Bing*** | ***MapQuest*** | ***OSM*** | ***HERE*** | ***Apple*** | ***TomTom*** | ***Yandex*** |
|---|---|---|---|---|---|---|---|---|
| *License* | Proprietary | Proprietary | Proprietary | **Open Database** | Proprietary | Proprietary | Proprietary | Proprietary |
| *Data Provider* | MAPIT, Tele Atlas, MDA Federal, DigitalGlobe, user contributions | NAVTEQ, NASA, Intermap, Blom, SK Planet, Ordnance Survey, Pictometry International | TomTom, OpenStreetMap, and others | **User contributions, open data and data donations** | NAVTEQ | TomTom, and others | TomTom, Tele Atlas | user contributions, NAVTEQ and others |
| *Map Type* | Map with traffic data, Satellite with Traffic Data, Hybrid | Road, Satellite, Hybrid, Bird's Eye, Traffic, 3D, London Street Map, Ordnance Survey Map, Venue Maps | Road, Satellite, Traffic | **Standard Map, Transport Map, Cycle Map, Humanitarian** | Map View, Satellite, Terrain, 3D, Traffic, Public Transportation, Heat Map, Map Creator, Explore Places, Community | Standard, Hybrid, Satellite. All include a traffic data layer | Standard, road, traffic, 3D | Standard, hybrid, satellite, traffic, 3D |
| *Street View* | Yes | Yes | No | **No** | No | No | No | No |
| *3D Model* | Yes (with plugin) Limited to certain areas | Yes (Windows 8/10) | No | **Yes, third-party** | Yes, limited to certain areas | Yes, limited to certain areas | Yes | Yes, limited to certain areas |
| *Live Traffic* | Yes | Yes (35 Countries) | Yes | **Yes, partial in a third-party** | Yes | Yes | Yes | Yes |
| *Weather* | No | | No | **Yes, third-party** | Yes | Yes | Yes | Yes |

vehicle signals only, if the knowledge of environment information is unknown. For example, a lane-change event can be represented by slight changes on steering angles or lateral accelerations. But when it takes place on a curved road, a vehicle could also move straight but enter another lane, even no steering angle changes appeared. In addition to signals captured for the driver and vehicle, environmental information from map data could be an additional source to benefit driving event recognition and assessment. As illustrated in Fig. 1, the first task of this study is to retrieve driving scenario attributes from map data, and combine with vehicle dynamic signals for driving event recognition. This task will be discussed in Sect. III and IV.

For autonomous driving perception, the effectiveness of sensor-based vision/laser systems is sometimes limited by illumination and whether conditions (e.g. shadow, snow, etc.). Prior knowledge of the surrounding scenario is expected to assist these systems and contribute to understand the driving scenario. To capture the environment information, one typical approach is Simultaneous Localization And Mapping (SLAM) [24, 25]. SLAM is usually based on laser sensors, which constructs and updates local environment by tracking a vehicle in the same environment multiple time. The scope of this paper is different. We are trying to take advantage of global available map data source and explore its potential and limits. Many studies have considered using map data to fuse with other signals for autonomous driving perception, vehicle localization, and road detection. Cao et al. [4] combined map data and camera images, improving GPS positioning accuracy to localize the vehicle at the lane-level. Wang et al. [5] integrated map data with image and Lidar signals, to provide a holistic 3D scene understanding on a single image, like object detection, vehicle pose estimation, semantic segmentation, etc. Alvarez et al. [6] considered using map data as prior knowledge, and investigated multiple image-processing methods for road detection. Seff et al. [7] retrieved road attributes from map data, hence providing a visual sense description. Laddha et al. [8] used the map data as supervised annotation, and designed deep learning approaches for road detection. The term of drivable space [9] is characterized as the static road surface restricted by what the moving object occupies, where the static geo-location and rough shape of roads can be retrieved from map data. In our study, the second task is to leverage map data, and explore how it can be integrated with other sensors for road perception [10]. Fig. 2 demonstrates the overall framework for the second task. The map data is employed to render virtual street views and prior road masks which will be further refined. Additional road masks are obtained from image processing and Lidar point clouding approaches. These masks will be fused to characterize the drivable road surface. The availability of using map data for deep learning will also be discussed. Image and Lidar processing methods are theory-driven, whereas deep learning approaches are data-driven. Therefore, one contribution of this work is to comprehensively explore the OSM capability in these two major aspects. Sect. V, VI, VII will cover this task.

The objective of this study is to explore the capability of using map data for environment understanding, including driving events recognition and road semantic segmentation. Therefore, we focus on how map data could be leveraged for these tasks. Even with limited experimental dataset, the performance is encouraging. This makes map data to be potentially used for more intelligent vehicle applications.

TABLE II
ENVIRONMENT ATTRIBUTES PARSED FROM OSM

| Group | Attributes | Data Type |
|---|---|---|
| *Direct* | One-way or Two-ways | Binary |
| | Number of lanes | Integer scalar |
| | Direction of each lane | Vector of integer |
| | Road type | Integer scalar |
| | Speed limit | Scalar |
| *Indirect* | Intersection type | Integer scalar |
| | At intersection | Binary |
| | Distance to intersection | Scalar |
| | Bearing angle to intersection | Scalar |
| | Road curvature | Scalar |
| | Heading angle | Scalar |
| | Distance to road center | Scalar |

## II. DATASET DESCRIPTION

### A. OpenStreetMap (OSM) Data

Table I compares several major available map resources. The OpenStreetMap (OSM) data [11] is selected in our probe study because it is open and free, which makes it widely used in research activities. However, the limit is that OSM data is provided based on user contributions, and therefore its accuracy is not always guaranteed. In general, a commercial used high-digits map data is expected to provide similar availability with higher precision.

The OSM data can be accessed via its website[1], and the data within an area of interest can be downloaded by specifying a bounding box in terms of latitudes and longitudes. The data is given in XML format and structured using three basic entities: nodes, ways, and relations [12, 13].

- *Nodes* – Nodes are point-shaped geometric elements which are used to represent a point-of-interest (POI), like traffic signs and intersections. A node is described by its GPS coordinates and a list of available tags. Each tag is formatted with a key and its corresponding value, describing the node attributes.
- *Ways* – Ways are used to model line-shaped or area-shaped geometric objects such as roads, railways, rivers, etc. A way entity is formatted as a collection of nodes.
- *Relations* – The relations are used to represent relationships between nodes and ways to form more complex structures.

### B. UTDrive Vehicle Dynamic Signals

Our previous studies [14] designed a smartphone application – Mobile-UTDrive, for use on in-vehicle signal capture. The device Inertial Measurement Unit (IMU) readings are converted to vehicle-referenced accelerations and rotations to represent vehicle dynamic signals. Previous studies also utilized a smartphone to simulate the touch-based and voice-based interface between the driver and the vehicle, and assessed driving performance from the vehicle dynamics feature space [15, 16].

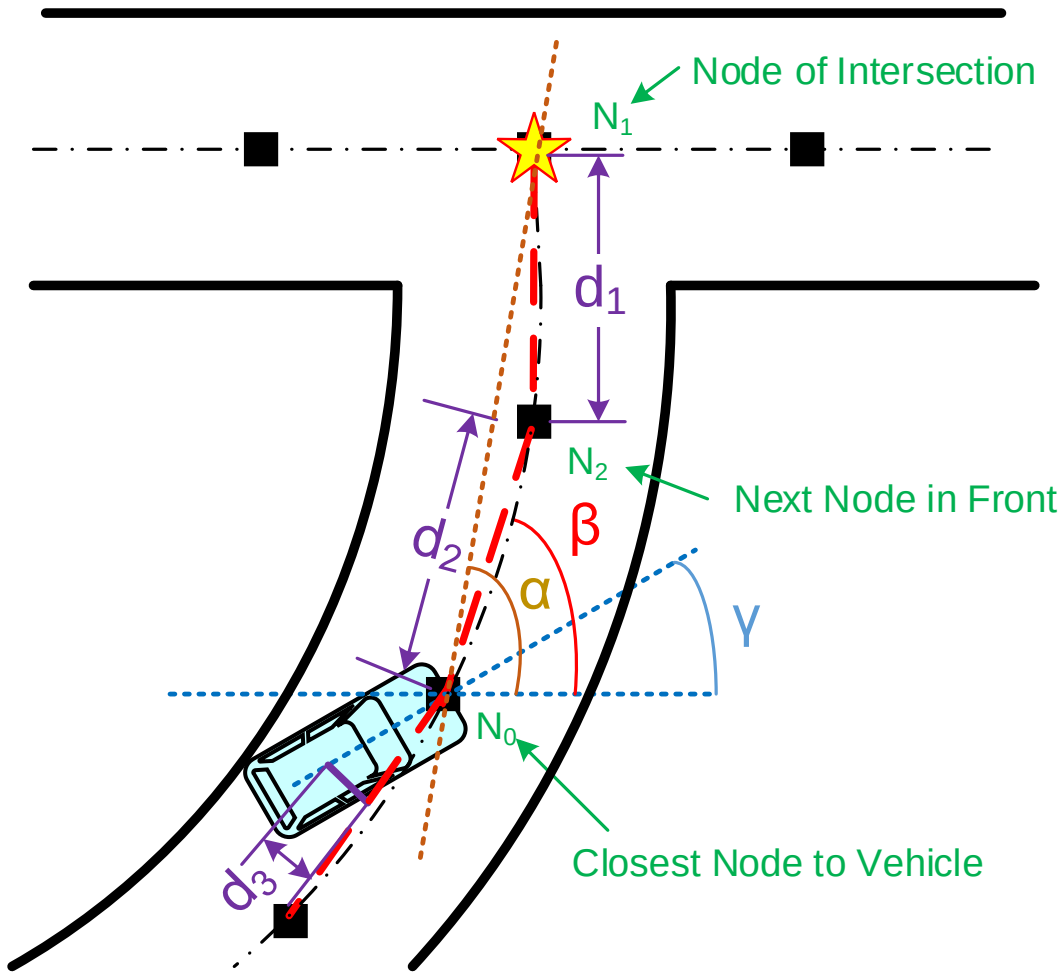

Fig. 3. Example of computational attributes. N0 is the closest node to vehicle, N1 is the node of intersection, N2 is the next node in front of driving direction. α is the bearing angle to the intersection, β is the road curvature, γ is the vehicle's heading angle. d1+d2 is the distance to the intersection, d3 is the distance to road center.

In this study, we continue to employ 3-axis accelerations and 3-axis rotations, as well as moving speed and heading angle collected from the GPS, to establish the vehicle dynamics feature vector for driving events recognition. Focusing on the steering control events specifically, driving events are manually partitioned as time-variant segments and labeled into five pre-set categories – Lane-Keeping (LK), Lane-Change-Left (LCL), Lane-Change-Right (LCR), Turning-Left (TNL), Turning-Right (TNR). The environment attributes retrieved from OSM will be added, to evaluate whether this knowledge could be useful.

### C. KITTI Road Benchmark

For the road segmentation experiment, the KITTI Road dataset[2] is introduced in this chapter [17]. KITTI is a designed computer vision benchmark suite for real-world autonomous driving perception challenges. The KITTI dataset has been recorded from a moving platform (Volkswagen Passat station wagon) while driving in and around the mid-sized city of Karlsruhe, Germany. The testbed includes two color camera images on the left and right side of the vehicle, one Velodyne 3D laser scanner, high-precision GPS measurements and IMU accelerations from a combined GPS/IMU system. The raw data has been further developed to provide stereo, optical flow, visual odometry, 3D object detection, 3D tracking, road detection, and semantic parsing sub-tasks.

In this study, the road detection benchmark package has been utilized, which consists of 289 training and 290 test images. The images are grouped into three different categories road scene:

- UU – urban unmarked (98 train / 100 test)
- UM – urban marked (95 train / 96 test)
- UMM – urban multiple marked lanes (96 train / 94 test)

Ground truth knowledge for training images has been generated by manual annotations and is available for two different road terrain types: (i) road – the road area, (i.e., the

[1] www.openstreetmap.org

[2] http://www.cvlibs.net/datasets/kitti/

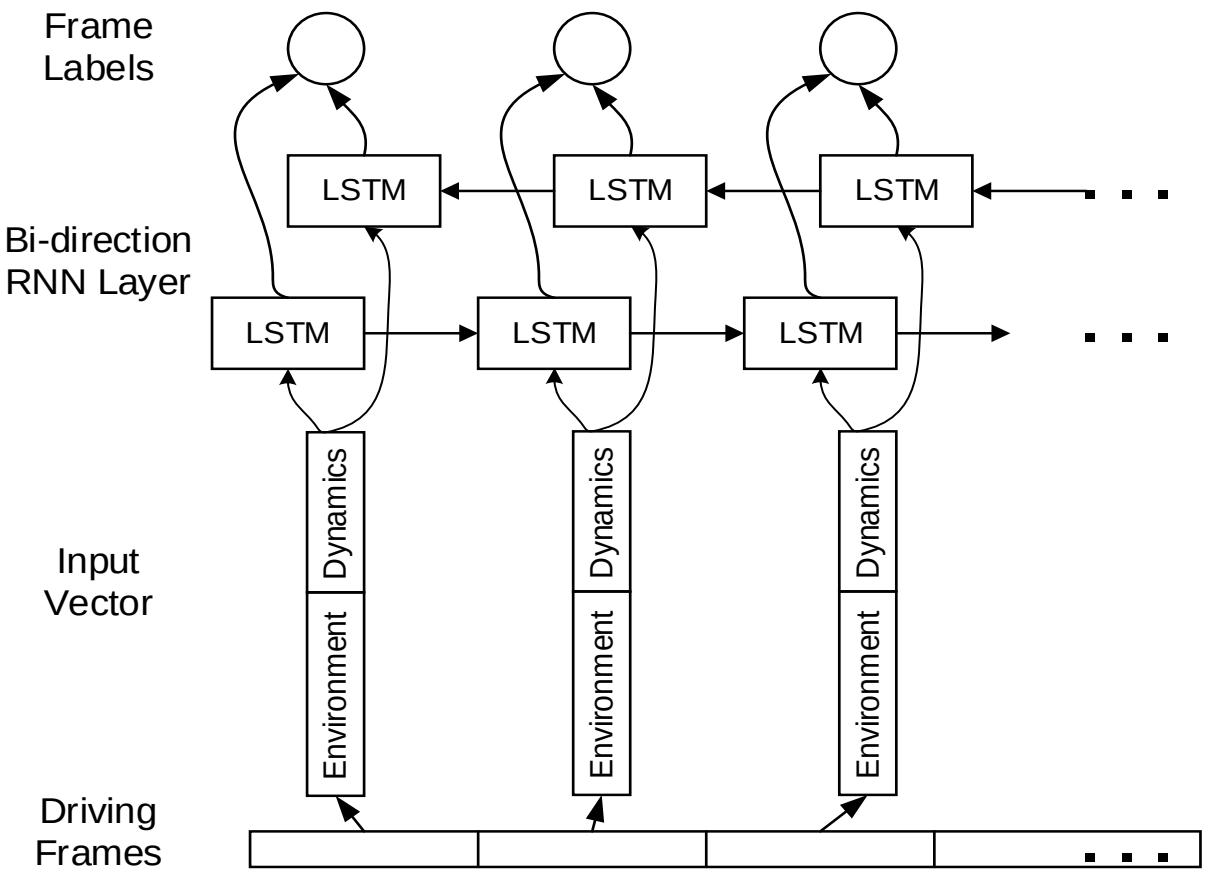

Fig. 4. Bi-RNN structure for driving event detection. The driving sequence data is segmented into frames, and the input vector contains both dynamic and environmental features. The output label is given on each frame.

TABLE III
DRIVING EVENT CLASSIFICATION CONFUSION MATRIX

(a) Without OSM Attributes

| | | Predicted | | | | | Acc. |
|---|---|---|---|---|---|---|---|
| | | *LK* | *LCL* | *LCR* | *TNL* | *TNR* | |
| Labeled | *LK* | **72** | 31 | 30 | 2 | 3 | 52.17% |
| | *LCL* | 3 | **37** | 5 | 0 | 0 | 82.22% |
| | *LCR* | 2 | 4 | **43** | 0 | 0 | 87.76% |
| | *TNL* | 2 | 0 | 0 | **22** | 1 | 88.00% |
| | *TNR* | 0 | 0 | 0 | 1 | **16** | 94.12% |

(b) With OSM Attributes

| | | Predicted | | | | | Acc. |
|---|---|---|---|---|---|---|---|
| | | *LK* | *LCL* | *LCR* | *TNL* | *TNR* | |
| Labeled | *LK* | **65** | 42 | 24 | 0 | 7 | 47.10% |
| | *LCL* | 2 | **43** | 2 | 0 | 0 | 95.56% |
| | *LCR* | 2 | 2 | **45** | 0 | 0 | 91.84% |
| | *TNL* | 0 | 1 | 0 | **23** | 1 | 92.00% |
| | *TNR* | 0 | 1 | 0 | 0 | **17** | 94.12% |

composition of all lanes), and (ii) lane – the ego-lane, (i.e., the lane the vehicle is currently driving on, only available for category "UM").

## III. ENVIRONMENT ATTRIBUTES RETRIEVAL FROM OSM

To retrieve environment attributes surrounding the vehicle, the current vehicle GPS location and OSM map data are required as input. The general idea is, given the vehicle GPS location, find the closest node within the map data, and parse or compute corresponding attributes associated with this node. The map data contains rich information, and desired attributes are selected according to specific tasks. Based on our application of driving event detection, a list of attributes has been decided which contains both (i) direct attributes and (ii) indirect attributes.

Table II lists the direct attributes that can be parsed from OSM data, as well as the indirect attributes that require geometry computation. The direct attributes are retrieved from the closest node tags. Since a node may be an element of either a way or other buildings, and the interest is to obtain road description information, the closest node should be selected within the way nodes only. For some explanation, the "direction of each lane" can be left only, left and through, through only, through and right, right only, left and right, and all directions. These seven types of directions are labeled as integers for convenience of computation. The road type is classified by OSM to be residential, tertiary, secondary, service, unclassified, and so on.

It is noted that a way is represented by a collection of nodes, and each node has its own GPS coordinates. Therefore, these way nodes can be referenced with the vehicle's GPS coordinates, and provide the vehicle's relative position on the road. Fig. 3 explains the computational attributes using an example, where $N_0$ is the closest node to the vehicle, and $N_1$ is the node of intersection, which is decided by the number of ways that are connected at the node. Based on the number of connected ways and the start/end point of each way, the type of intersection is classified as either: crossing, T-junction, turning, merge, and exit. Because of the road curvature, the distance to the intersection is not exactly the distance between vehicle node $N_0$ and intersection node $N_1$, but the sum of way segments partitioned by nodes ($d_1$+$d_2$). When the GPS coordinates are projected onto the Universal Transverse Mercator (UTM) flat plane frame, the bearing angle $\alpha$ is computed as the east-based direction between $N_1$ and $N_0$; the road curvature $\beta$ is computed between $N_0$ and the next node in front $N_2$; and the vehicle's heading angle $\gamma$ is given by its GPS data package. Since the OSM way nodes are defined at the way center, the distance to the road center $d_3$ is measured by the perpendicular distance from the vehicle's location to the way segment line.

## IV. OSM FOR DRIVING EVENT RECOGNITION

The retrieved environment attributes will be submitted to combine with vehicle dynamic signals for driving event recognition. In this section, we take advantage of deep learning approaches, to segment and classify driving sequence data into: Lane-Keeping (LK), Lane-Change-Left (LCL), Lane-Change-Right (LCR), Turning-Left (TNL), and Turning-Right (TNR).

To recognize driving events within a long driving session, our previous studies [1, 27] tried to capture vehicle dynamic changes in time series, and use these features to specify event segments. The nature of this task is to assign labels to each entity within a sequence, which is close to the sequence-tagging problem seen in Natural Language Processing (NLP) area for sentence analysis. We adopt a similar model given by [18] and [19], which is based on a Bidirectional Recurrent Neural Network (Bi-RNN) with Long Short-Term Memory (LSTM) units. In NLP, Bi-RNN and LSTM have been shown successful to carry both long-range dependencies and short-range attentions to understand a text sentence, and output a sequence of labels on each word to represent its contextual meaning. Similarly, a long driving sequence is composed of frames of driving events. Vehicle dynamics and environment information within each event, as well as their changes associated with neighboring frames should be considered in both short and long ranges to understand the driving task. The word embedding vector input in Bi-RNN are replaced by a list of vehicle dynamic features and environment attributes, and their word contextual representation outputs will correspond with our driving event labels. In naturalistic driving, lane-keeping events occur for most of the time duration while other events have fewer occurrences. In the cost function design, if all five event classes are considered uniformly, the system will emphasize LK and therefore limit the effectiveness of lane-change and turning events. To reduce this imbalance effect, we modify the cost function by reducing the weighting factor of LK (e.g., by half) on the top layer, and therefore the final cost function $C_{final}$ is defined as:

$$C_{final} = C_{LCL} + C_{LCR} + C_{TNL} + C_{TNR} + 0.5 \cdot C_{LK} \qquad (1)$$

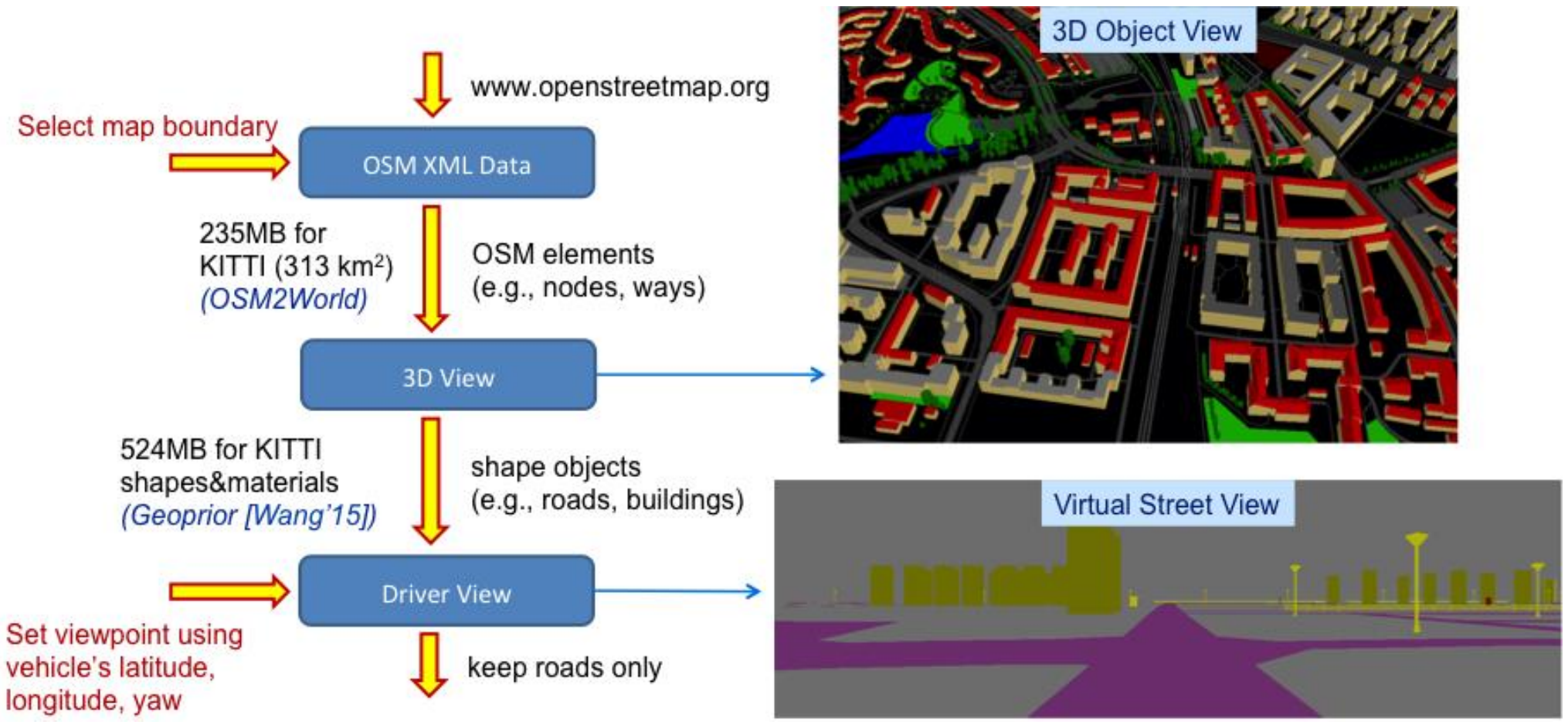

Fig. 5. Rendering virtual street view from OSM, with approach flow diagram and output example.

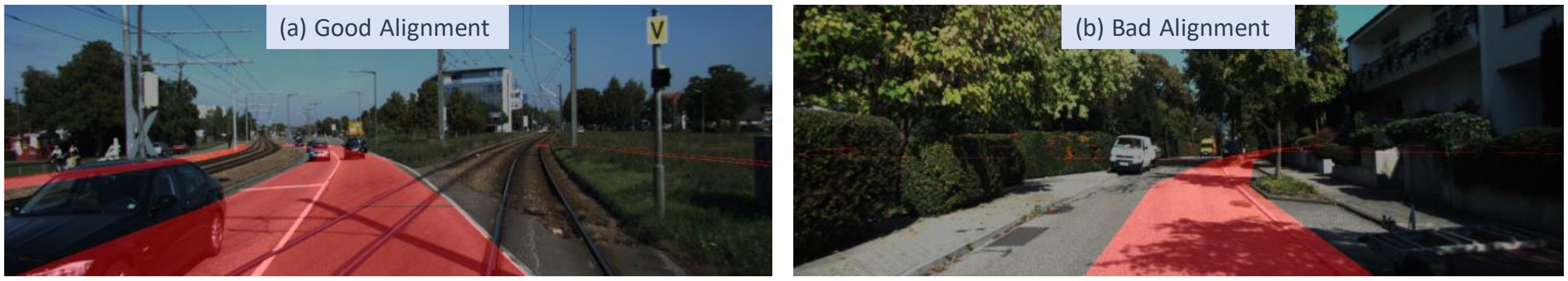

Fig. 6. The effect of overlaying the OSM road mask (displayed in red color) onto the real camera image. (a) good alignment and (b) bad alignment.

The downside of this modification is that it will increase false-positives for lane-change and turning events. This is reasonable if the goal is to reduce the miss rate of these events (but allow a small false alarm). The weighting factor can be adjusted according to other tasks.

The model architecture is illustrated in Fig. 4. A long sequence driving data is segmented into fixed window frames (e.g., 0.1 second). The vehicle dynamic signals (accelerations, rotations, moving speed, and heading angle) and the retrieved environment attributes (Table II) are concatenated to form the input vector. The hidden layer is established in a Bi-RNN structure with LSTM units. An output event label is assigned to each frame, and supervised ground truths are given by human annotation. There are 102 driving data sequences for the experiment, and we use 65 sequences for training and the remaining 37 for test. The number of training epoch is set to 80, and each mini-batch contains 4 sequences. The average sequence length is about 1 minute each, and we select 700 frames (i.e., 70 second) as the maximum sequence length with zero padding. The input vector dimension is 25, and the hidden layer unit size is 300. A 25% dropout is applied to reduce over-fitting in training. The learning rate is initialized as 0.001 with a decay rate at 0.9. These hyper-parameters are selected by trial-and-error, and finalized with those obtained the best results.

In the test dataset, we have 138 LK, 45 LCL, 49 LCR, 25 TNL, and 17 TNR driving events. The recognition effectiveness is evaluated using a 5x5 confusion matrix for the five events, shown in Table III. Since the event classification label is given on each frame, there may be a discontinuity between adjacent labels. The evaluation is therefore conducted on chunks, which are segmented by the ground truth boundaries. Within each chunk, the largest predicted label will represent the label of this chunk. Table III-(a) shows the results without OSM attributes and Table III-(b) shows those with OSM attributes. By adding the OSM environmental attributes, the lane-change detection accuracies have increased from 82.22% to 95.56% for LCL, and 87.76% to 91.84% for LCR. The road curvature, heading angle, distance to road center, and other lane information, make contributions for the improvement. For TNL and TNR events, the distance to intersection is also a useful feature. However, the LK accuracy decreases, and false-positives are increased for lane-change events. This is because lane-change and lane-keeping events usually take place on the same road, where they share common road information, which makes them even more difficult to distinguish.

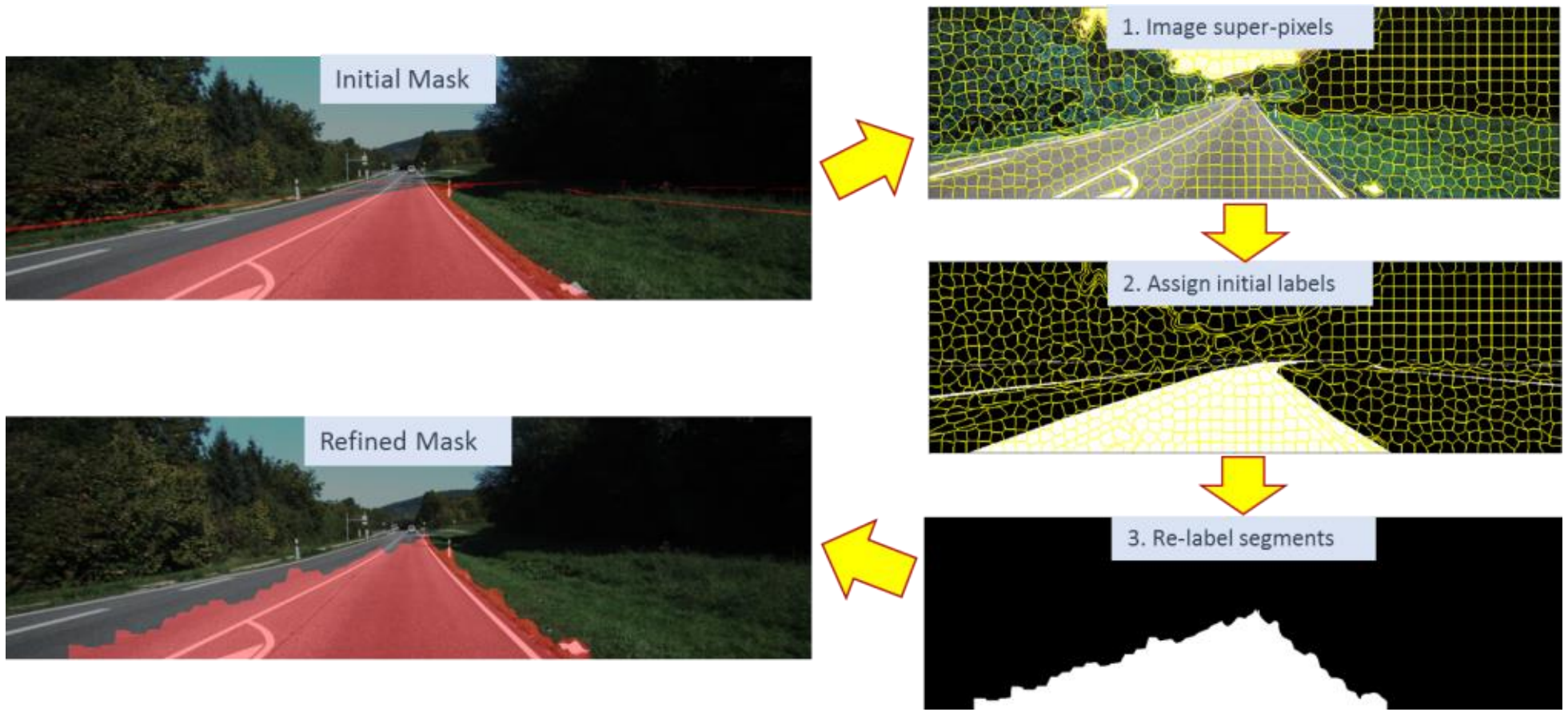

Fig. 7. Super-pixel relabel approach for the road mask refinement, with processing steps and result example.

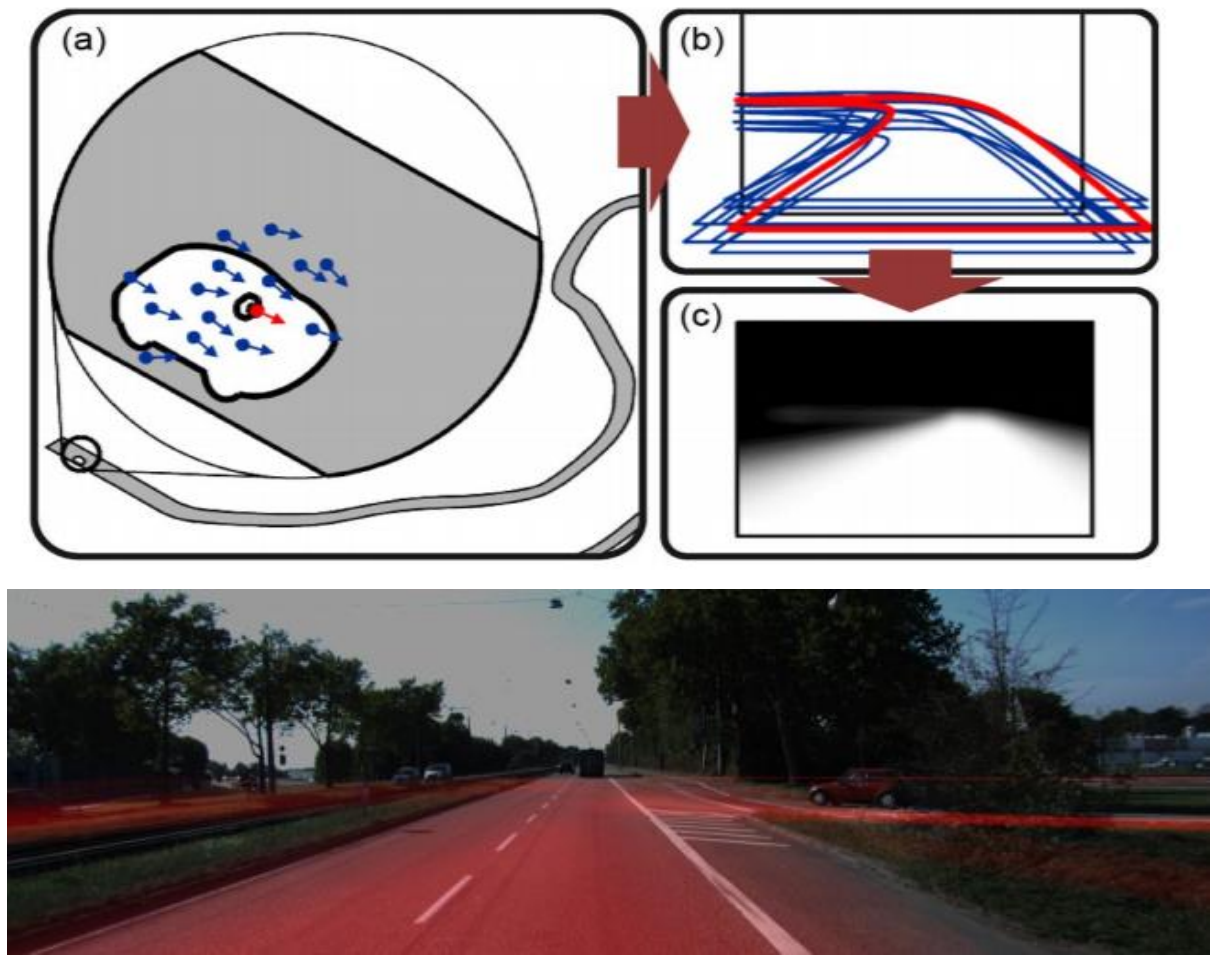

Fig. 8. Multiple candidates approach to generate a confidence road mask, with processing steps and result example.

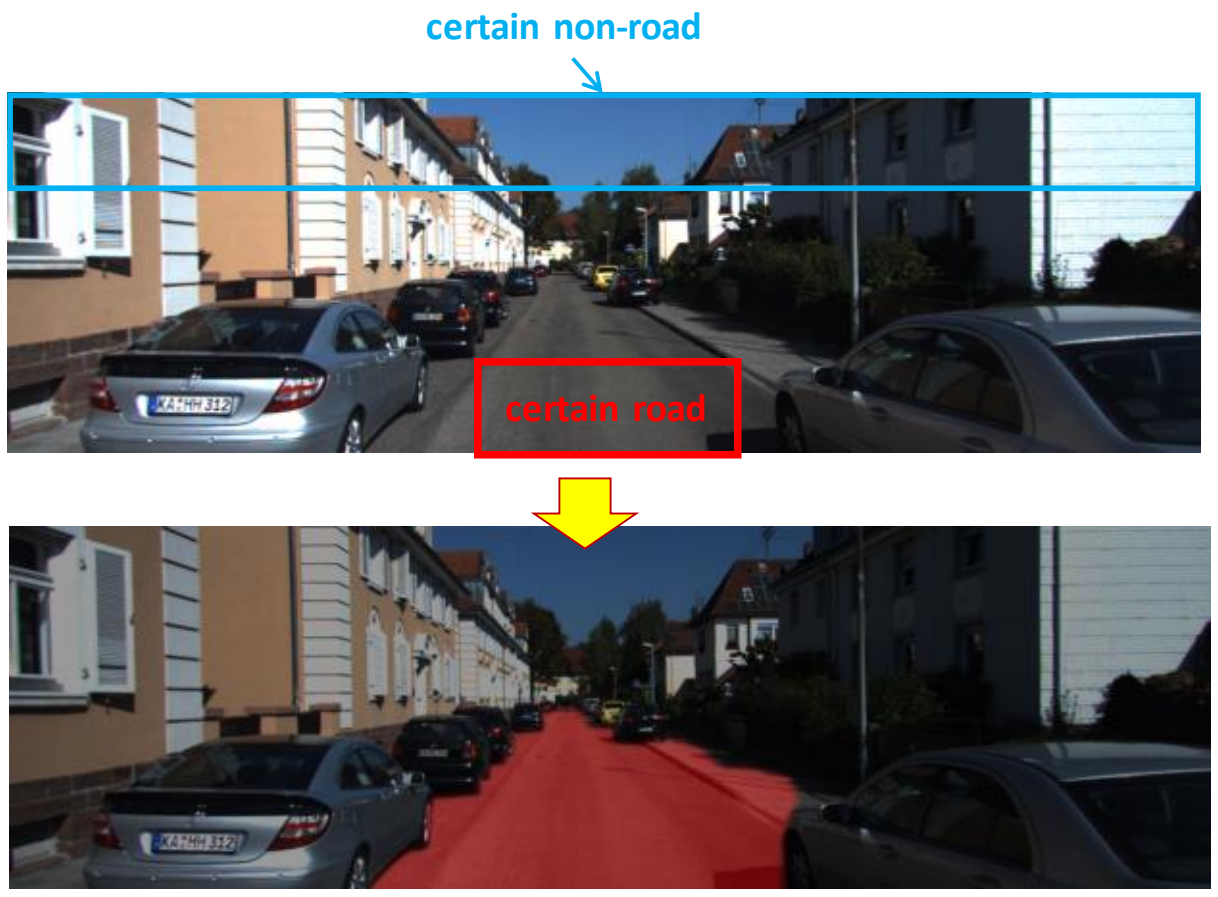

Fig. 9. Example of GrabCut algorithm. Two hand-labeled rectangles are given on an image, and the algorithm outputs road and non-road segments.

## V. Virtual Street View Rendering from OSM

Fig. 5 illustrates the general approach of virtual street view rendering, as well as two key output examples from middle steps. As stated in Sect. II, the OSM data can be downloaded from its website by specifying a bounding box in terms of latitudes and longitudes. The data is structured in XML format, which describes node, way, and relation elements. The OSM2World toolkit[3] can be used to create 3D models from the OSM XML data. It renders a 3D virtual world from the birds-eye view, and generates a shape object (e.g. road, building, etc.) description file. Given the vehicle's GPS coordinates and its heading direction, we can set it as a viewpoint and project the 3D virtual world into the driver's perspective [5]. If the interest is to generate the road mask only, we keep road objects and remove others, and generate the virtual street view as a binary road mask.

The KITTI road dataset contains real images, Lidar point clouds, as well as GPS and IMU information, which makes it possible for us to compare the virtual street view against the real-world scenario. Fig. 6 displays two examples that contain OSM road masks in the virtual street view overlaid onto real camera images. One advantage of the OSM road mask is that, it can display road surface area visually hidden by buildings or trees, and therefore provides more information on the road trajectory. There is a good alignment in Fig. 6-(a), and bad alignment in Fig. 6-(b). In the bad alignment, there is a shift between the virtual mask and the real road, which is caused by a GPS error; and a mismatch on the road width, which is caused by an OSM error. Since the OSM data is provided by user contributions, the OSM accuracy cannot be guaranteed. Therefore, more processing steps are needed to reduce the misalignment.

### A. *Super-pixel Refinement Approach*

One refinement approach is to consider in super-pixels [8]. The processing steps are shown in Fig. 7. Super-pixel segmentation is based on the real camera image, using a K-means clustering method. Specifically, an image size is 375*1242 pixels, and we select K=800 to generate 800 super-pixel segments. Initial pixel-wise labels from the OSM road mask are overlaid onto the real image and assigned to all super-pixel segments. If a segment contains more than 50% initial road labels, it should be relabeled as a road super-pixel, and non-road otherwise. Once all super-pixels are relabeled, the largest connected component will be selected as the refined road mask. It is worth noting that this approach has been shown to clip the boundary of the initial road mask; it removes some boundary coverage and therefore may result in more false-negatives.

[3] http://osm2world.org/

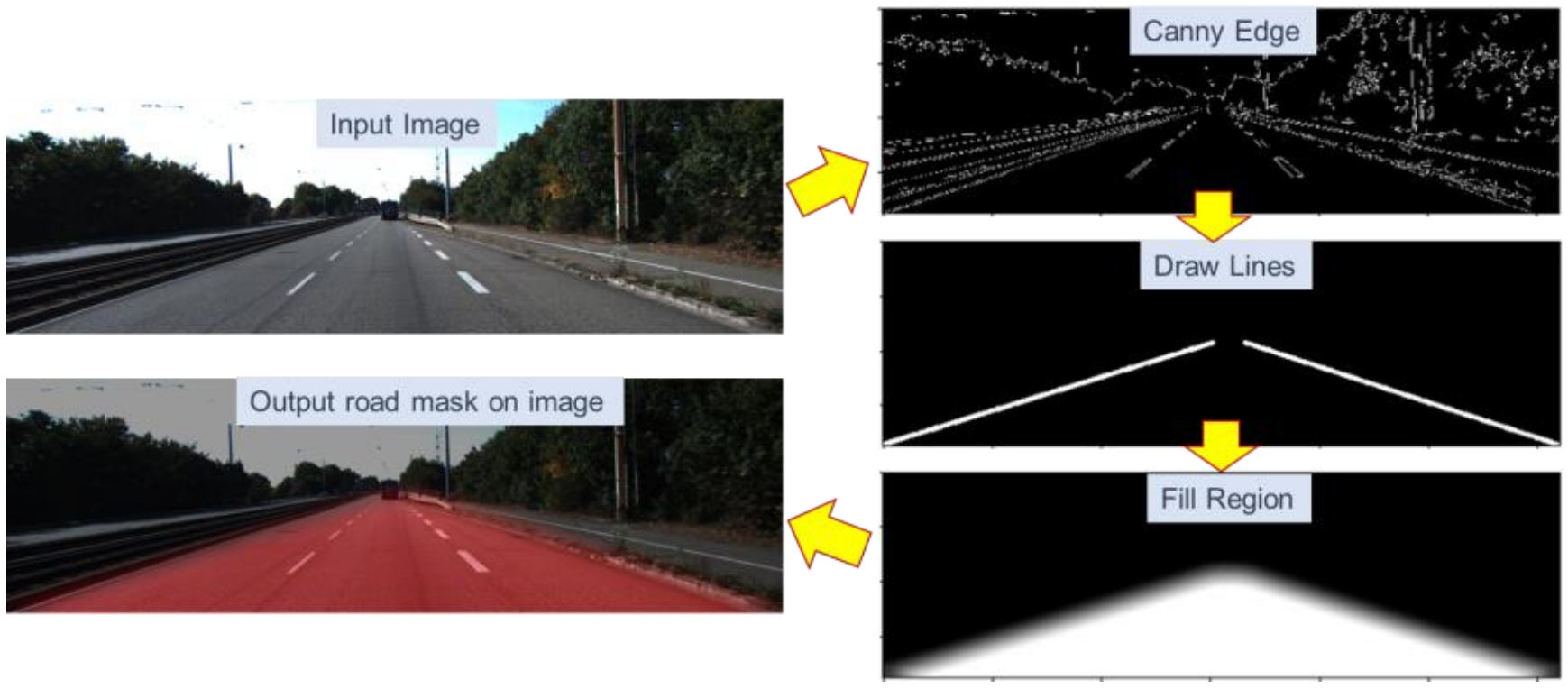

Fig. 10. Lane-mark detection approach for road segmentation, with processing steps and result example.

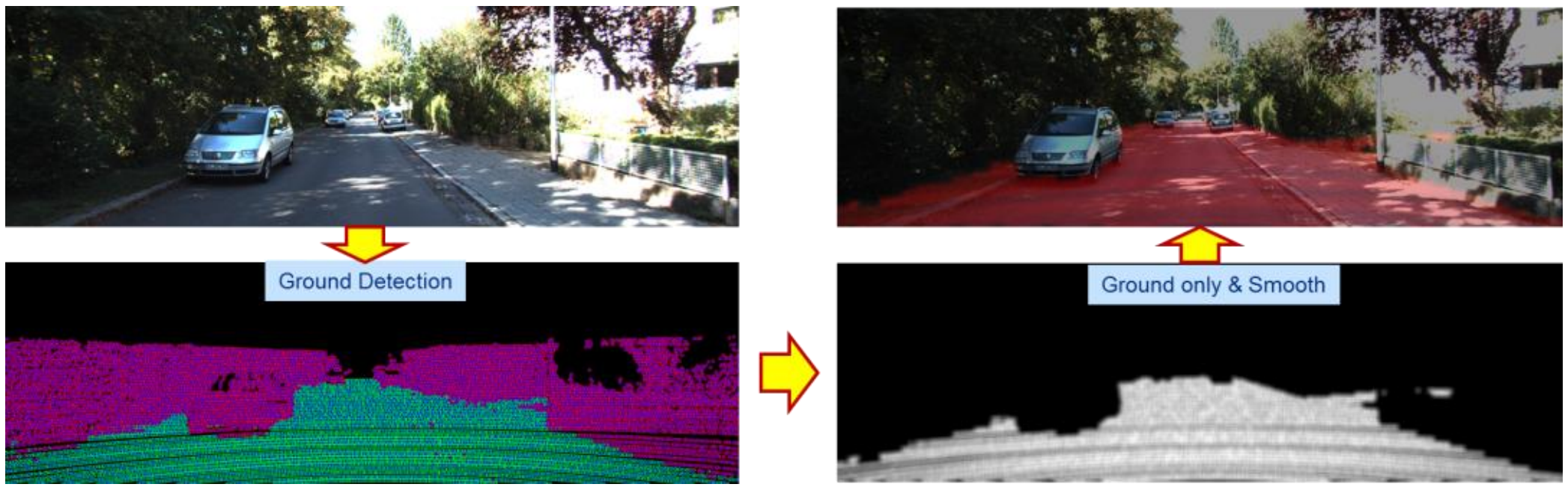

Fig. 11. Lidar point cloud processing for ground detection, with processing steps and result example.

### B. Multiple Candidates Approach

The super-pixel refinement output is still a binary road mask, where an alternative multiple candidates approach [4, 6] will create a confidence road mask. The processing steps are shown in Fig. 8. When the original vehicle GPS location and moving direction is projected onto the UTM flat plane, the corresponding x, y coordinates and perspective angle is obtained. Considering the GPS error, a more precise position candidate could be located nearby, within the GPS variance range (e.g., x and y within ±1 meter, and angle within ±10 degrees). Therefore, N (e.g., N=100) possible viewpoint candidates are randomly selected within the variance, based on a uniform or Gaussian distribution. Virtual street views will therefore be rendered at the N viewpoint candidates, generating N binary road masks. The final confidence mask will be the overlay of all candidate road masks, with an average on each pixel given by,

$$P(x_i) = \frac{1}{N}\sum_{i=1}^{N} B_j(x_i) \tag{2}$$

where $x_i$ denotes the $i$-th pixel in the image, $B_j$ denotes the $j$-th candidate binary mask, $N$ denotes the number of candidates, and $P(x_i)$ denotes the probability for $x_i$ being the road. In contrast to the super-pixel refinement approach, the multiple candidate approach brings in greater coverage of the road area, and therefore will result in more false-positives.

## VI. OSM for Sensor Fusion

Admittedly, the OSM data can only provide apriori knowledge of the static driving scenario, and the rendered virtual street view can only generate a coarse road mask. A more precise characterization of the drivable space would still rely on vision/laser sensors. This section will introduce two image processing methods and one Lidar point cloud processing method, and discuss how the OSM prior knowledge can be fused with these sensors.

### A. GrabCut Algorithm

The GrabCut algorithm was designed by Rother et al. [20], which is an interactive approach to extract foreground and background segmentations on an image. This approach requires a manually selected small area for the certain foreground region, as well as another small area for the certain background region. Based on these initial labels, two Gaussian Mixture Models are applied to model the foreground and background, and create the pixel distribution in terms of color statistics. All other unlabeled pixels will be assigned with the probability of being foreground or background. Next, a graph with a source node (connecting to the foreground) and a sink node (connecting to the background) is built from the pixel distribution, in which the other nodes representing the pixels and the weighted edges represent the probabilities. Next, a min-cut algorithm is used to segment the graph, partitioning all foreground pixels connected to the source node and all background pixels connected to the sink node. The process is continued until the classification converges or a pre-defined iteration step is reached. Fig. 9 shows an example of how the GrabCut algorithm is utilized for road detection. In our case, the top-bar is pre-selected as the certain non-road background, and the mid-bottom area is pre-selected as the certain road foreground. The GrabCut algorithm will start processing on these two rectangles, and segment the entire image into road and non-road regions. It is worthwhile to note that the GrabCut algorithm is based on color, and the effectiveness is limited if there is shadow or bad illumination conditions.

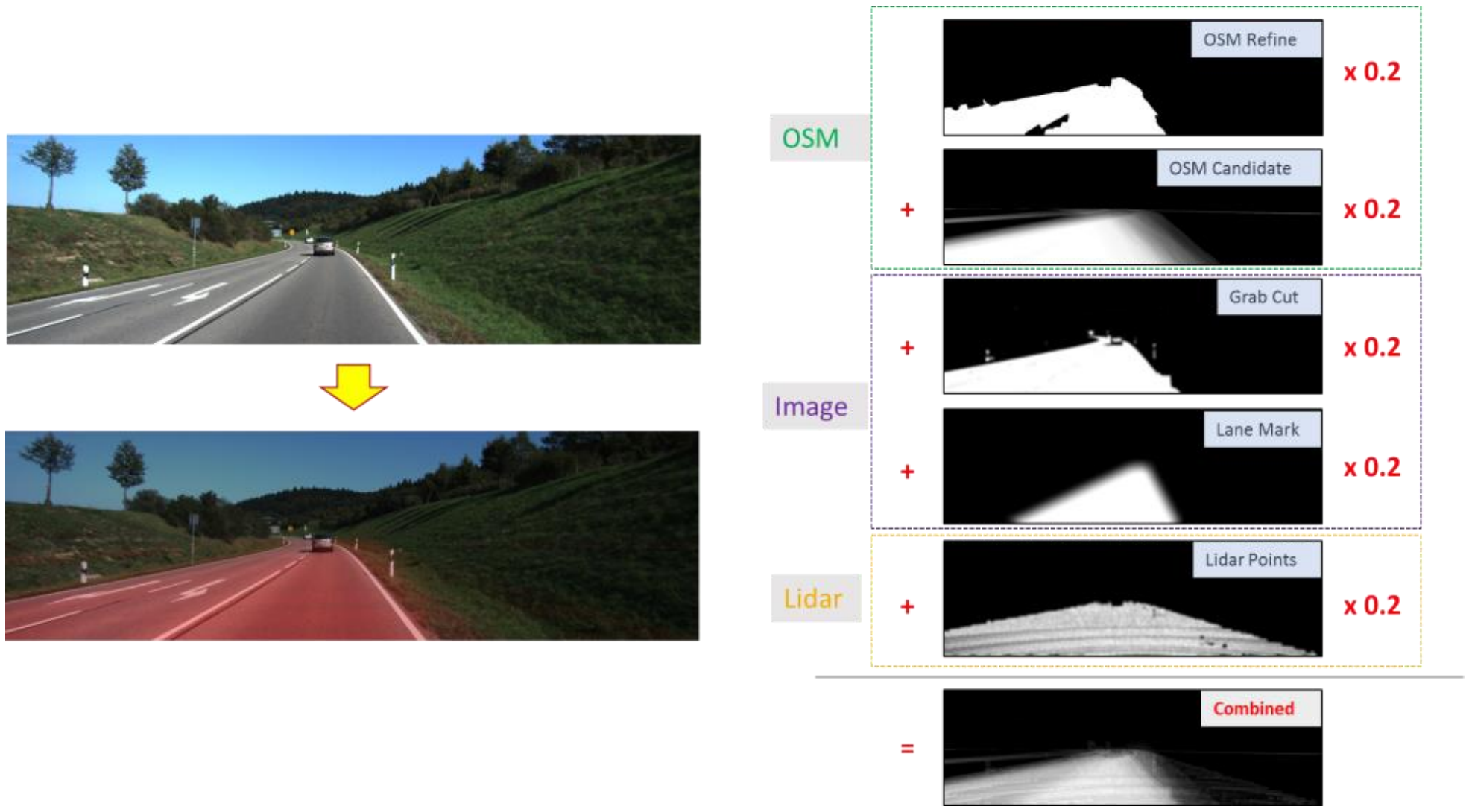

Fig. 12. The combined road mask is obtained by a weighted sum up of the 5 masks. The weighting factor in this example is selected 0.2, to consider the 5 masks evenly.

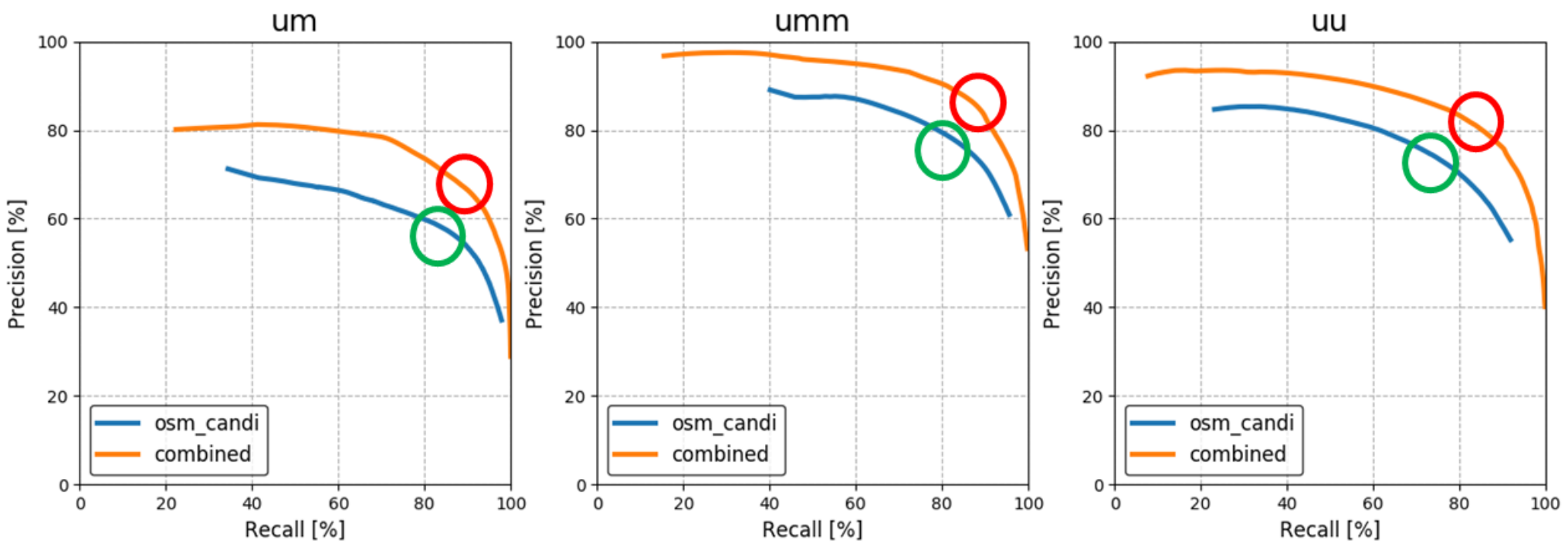

Fig. 13. Precision-recall trade-off curves for the confidence masks generated by the OSM multiple candidates approach and the combined approach. The circled locations indicate the selected thresholds, corresponding to the results shown in Table IV.

### B. *Lane-Mark Detection*

Another widely used image processing method is to locate lane-marks [21]. This approach is based on edges. Fig. 10 shows an example of these processing steps. First, a Canny edge detection is applied, based on the color gradients. Next, a Hough line transformation technique is utilized to draw lines with polynomial curve fitting. Finally, the region between the lane marks is filled as the road label. The advantage of this lane-mark detection approach is its robustness to image noise, however, it assumes that the lane marks are straight or almost straight, and therefore limits its effectiveness when there is a sharp-curved road or the lane boundaries are not clear.

### C. *Lidar Point Cloud Processing*

Since the KITTI dataset provides Velodyne Lidar point clouds, Lidar-to-image calibration matrices and projection matrices, it is possible to project the points onto the image plane to view the points from the driver's perspective. Next, a multi-plane technique [22] is utilized to segment the ground plane against other buildings and objects. The points classified as ground are kept and others are removed. Due to the points resolution, they may look sparse on the image, and therefore a smoothing step is added to generate the filled road mask. An example of the processing steps is shown in Fig. 11. Since the ground detection is based on the height computation of the point-cloud planes, it is difficult to separate the road surface against the sidewalk or nearby grass.

### D. *Sensor Fusion*

From the aforementioned approaches, five types of road mask are generated. Two masks from OSM (super-pixel refinement and multiple candidates), two from image processing (GrabCut algorithm and lane-mark approach), and one from Lidar point cloud processing. A combined road mask can be obtained by a weighted sum of all these masks. Fig. 12 displays an example for the combined mask overlaid on the image, in which the individual weighting factor is chosen as 0.2 to consider all five masks evenly. The weighting factors can be adjusted by how much confidence each approach is gained, or consider the weight selection as an optimization problem to calculate. Since the purpose of this study is to demonstrate that OSM could provide the road prior knowledge and could be considered together other sensor measurements, we did not expand the discussion on the weights selection. In addition, if more masks were obtained by other approaches (e.g. depth layout), they can also be added into the combined mask.

TABLE IV
ROAD DETECTION RESULTS USING OSM, IMAGES AND LIDAR MASKS

| | urban marked lane (UM) | | | urban marked multi-lanes (UMM) | | | urban unmarked lane (UU) | | |
|---|---|---|---|---|---|---|---|---|---|
| | *Precision* | *Recall* | *F1-Score* | *Precision* | *Recall* | *F1-Score* | *Precision* | *Recall* | *F1-Score* |
| **OSM direct** | 0.5123 | 0.8738 | 0.6177 | 0.7404 | 0.8023 | 0.7575 | 0.6689 | 0.6282 | 0.6319 |
| **OSM refinement** | **0.5684** | 0.8662 | 0.6555 | **0.7974** | 0.7713 | 0.7675 | **0.7054** | 0.6068 | 0.6368 |
| **OSM candidate** | 0.5516 | **0.8833** | 0.6441 | 0.7795 | **0.8258** | 0.7925 | 0.7815 | **0.6598** | 0.6952 |
| **Image GrabCut** | 0.6170 | 0.9300 | **0.7135** | 0.8724 | 0.8346 | **0.8369** | 0.7644 | 0.8600 | **0.7889** |
| **Image LaneMark** | **0.6032** | 0.8476 | 0.6722 | **0.8475** | 0.6167 | 0.7006 | 0.6602 | **0.8096** | 0.7114 |
| **Lidar PointCloud** | **0.3334** | 0.9885 | 0.4840 | **0.5959** | 0.9705 | 0.7340 | **0.4670** | 0.9113 | 0.5957 |
| **Combined** | **0.6396** | **0.9253** | **0.7293** | **0.8962** | **0.8207** | **0.8504** | **0.8193** | **0.8169** | **0.8089** |

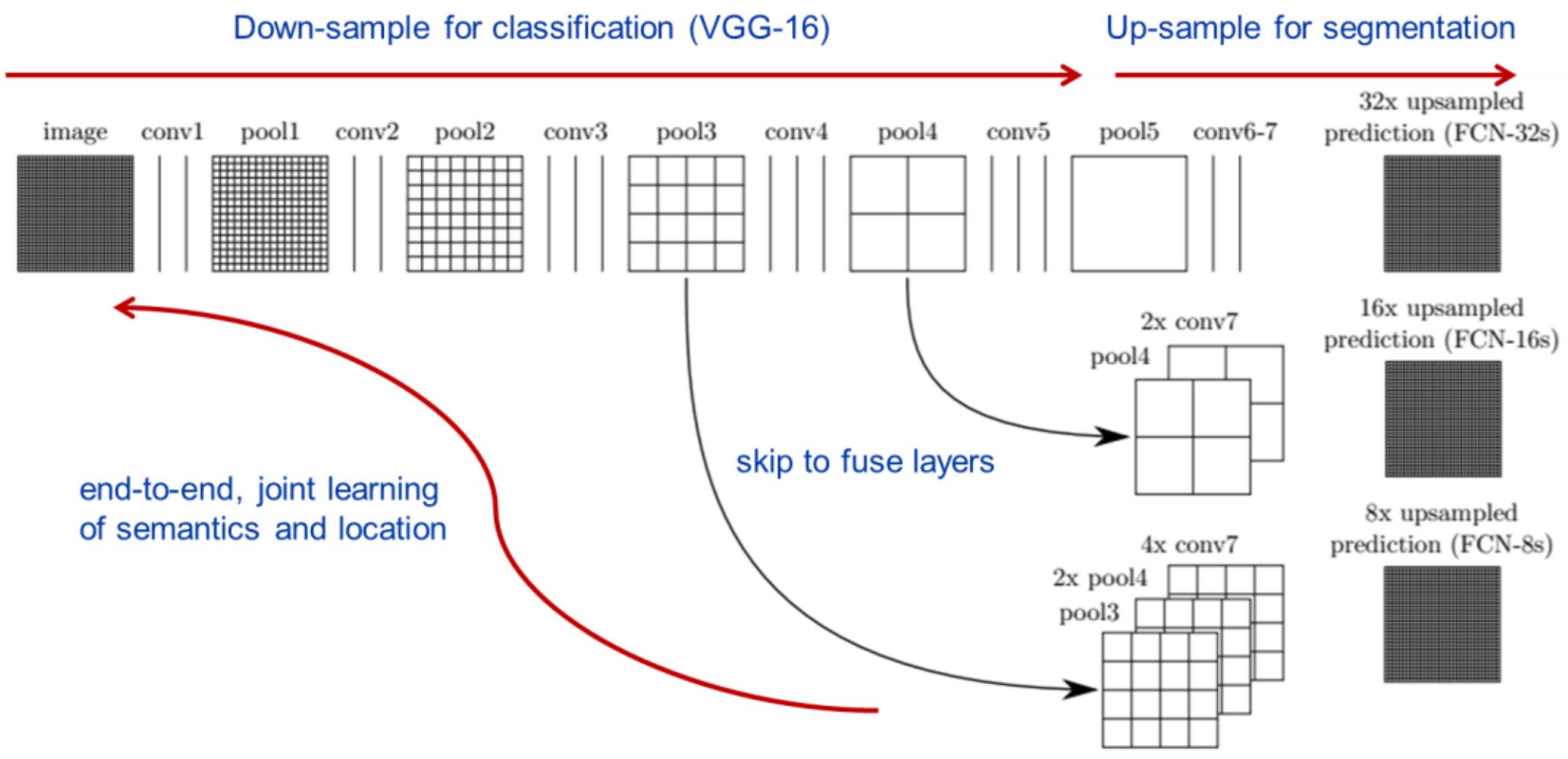

Fig. 14. FCNN architecture for image semantic segmentation.

Table IV summarizes the road detection results using the masks generated from OSM, images, and Lidar. The experiments are examined using the three road types in KITTI – urban marked lane (UM), urban marked multi-lane (UMM), and urban unmarked lane (UU). The quantity evaluation is conducted on each pixel, and results are measured using precision, recall and F1-score. Since the OSM multiple candidates approach and combined output are represented as confidence masks, a threshold should be selected to balance the precision-recall trade-off. The listed results are obtained by selecting the circled location on the trade-off curves in Fig. 13. Comparing the second and third row against the first row in Table IV, the OSM super-pixel refinement approach increases precision from the OSM direct rendering mask, and the OSM multiple candidates approach increases recall. This approves our discussion in Sect. V that the super-pixel refinement approach removes the boundary area while the multiple candidates approach brings in new coverage. Although the Lidar point cloud processing achieves high recall values, its precision values are the lowest because of the bicycle lane, sidewalk, and grass areas included. The image processing GrabCut algorithm results in higher F1-score than other individual masks. The best result is obtained from the combined mask, in which a uniform weighting factor (i.e., 0.2) is applied on all five masks. Since the GrabCut algorithm provides the best individual result, it is believed that a higher combined mask result is possible if a higher weight is selected on the GrabCut mask and lower weights on others.

## VII. OSM FOR DEEP LEARNING

Recently, deep learning approaches have been proven to be successful in computer vision area and autonomous driving applications. Deep learning does not require hand-crafted features, which is able to learn more information than traditional image processing approaches. The road prior knowledge obtained from OSM is an additional source isolated from the image itself, and therefore it is of interest to see how OSM can contribute to the road detection task with deep learning.

Road detection is a semantic segmentation task, which can be addressed by a widely used Fully Convolutional Neural Network (FCNN) [23]. Fig. 14 illustrates the FCNN architecture, which is composed of a VGG-16 convolutional network in lower layers for down-sampling and classification, and an up-sampling layer on the top to recover the image to its original size. Overall, it is an end-to-end jointly learning framework to capture image semantics at its locations.

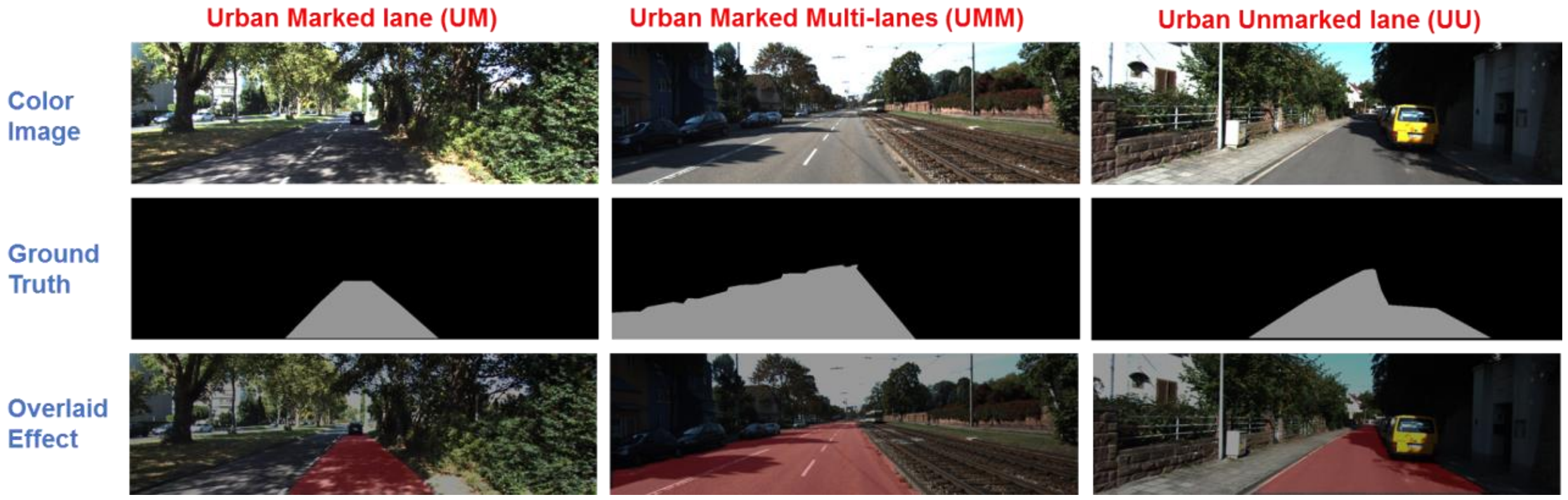

Fig. 15. Comparison for different lane type scenarios. The lane type categories are grouped into three columns, which are UM, UMM, and UU from left to right. The first row shows the color image, the second row shows the ground truth road mask given on each image, and the third row shows the overlaid effect with mask on each image.

TABLE V
ROAD DETECTION RESULTS USING DEEP LEARNING

| | **urban marked lane (UM)** | | | **urban marked multi-lanes (UMM)** | | | **urban unmarked lane (UU)** | | |
|---|---|---|---|---|---|---|---|---|---|
| | *Precision* | *Recall* | *F1-Score* | *Precision* | *Recall* | *F1-Score* | *Precision* | *Recall* | *F1-Score* |
| **Image only** | 0.8101 | 0.9383 | 0.8492 | 0.9267 | **0.8717** | **0.8858** | 0.8490 | 0.9282 | 0.8768 |
| **OSM only** | 0.7984 | 0.9391 | 0.8418 | 0.9040 | 0.8571 | 0.8672 | 0.8538 | 0.9281 | 0.8800 |
| **Image + OSM** | 0.8044 | 0.9478 | 0.8475 | 0.9291 | 0.8605 | 0.8787 | 0.8361 | 0.9265 | 0.8680 |
| **Mask only** | 0.7665 | **0.9492** | 0.8251 | 0.9321 | 0.8454 | 0.8712 | 0.8620 | 0.9362 | 0.8896 |
| **Image + mask** | **0.8281** | 0.9387 | **0.8648** | **0.9383** | 0.8510 | 0.8750 | **0.8664** | **0.9405** | **0.8922** |

To integrate OSM prior information into the FCNN architecture, one typical approach is to use the rendered virtual road mask as an isolated image, or as an additional channel combining with an RGB color image. In summary, this section compares five different types of input images:

- Camera image only – raw color image given by KITTI.
- OSM mask only – using the confidence mask obtained by OSM multiple candidates approach.
- Image + OSM – adding the OSM mask as an addition channel to the color image.
- Combined mask only – using the overlaid mask obtained by combining OSM, image processing, and Lidar processing approaches.
- Image + Combined – adding the combined mask as an additional channel to the color image.

It is worthwhile to note the diversity/inconsistency on the given ground truth. KITTI provides 289 annotated images, where the size is not big, but these images are categorized into three scenarios. As shown in Fig. 15, the UM ground truth is given only to cover the current ego lane, whereas the UMM and UU ground truth cover all the road surface. This will provide different supervision for the model to learn. However, due to the limited data size, we decide not to distinguish the categories in the training phase, but compare the differences on the test results.

Table V shows the road detection results using the deep learning approach, comparing variant inputs for the same FCNN model structure. The highest values in each column are highlighted, which are always greater than individual masks or fused results shown in Table IV. The most obvious finding is, the UM result is always lower than UMM and UU, producing higher recall but lower precision. This is because the UM ground truth only covers the current ego lane, which is smaller than the predicted road surface. Depending on the task interest, we should decide how to supervise the model training. However, the rendered OSM road masks cover all the road surface for the three categories. Even if the OSM node tags provide complete road attributes to decide the number of lanes, but without sufficient GPS accuracy, it is still challenging to locate the vehicle at the lane-level. This could be the limit for the current available OSM data.

In Table V, no matter what kind of input is given, image, mask, or any combination, the performance difference is not significant. This may possibly be explained in Fig. 16, which compares the training loss history for the five input types. If sufficient training iteration is given, any of the five input types will finally converge. The only difference is their convergence speed. The "mask_only" input converges the fastest, because it is already composed of high-level features extracted by image processing. For the "OSM_only" input, it does not contain as much information as the color image, so it takes more time for the model to learn, but can still converge in the end. This motivates several other potential capabilities to use the OSM data in deep learning for special cases.

### 1) *Special Case (a): Use OSM virtual street views for test*

There may be several pre-trained models published for road segmentation. Their models were trained using the camera street images and human annotation for the ground truth. We expect to employ their models in our experiments, but it is possible that the street view images are not always available in our testing area. Therefore, can we use the virtual street views rendered from OSM for testing?

*2) Special Case (b): Use OSM road mask for automatic annotation*

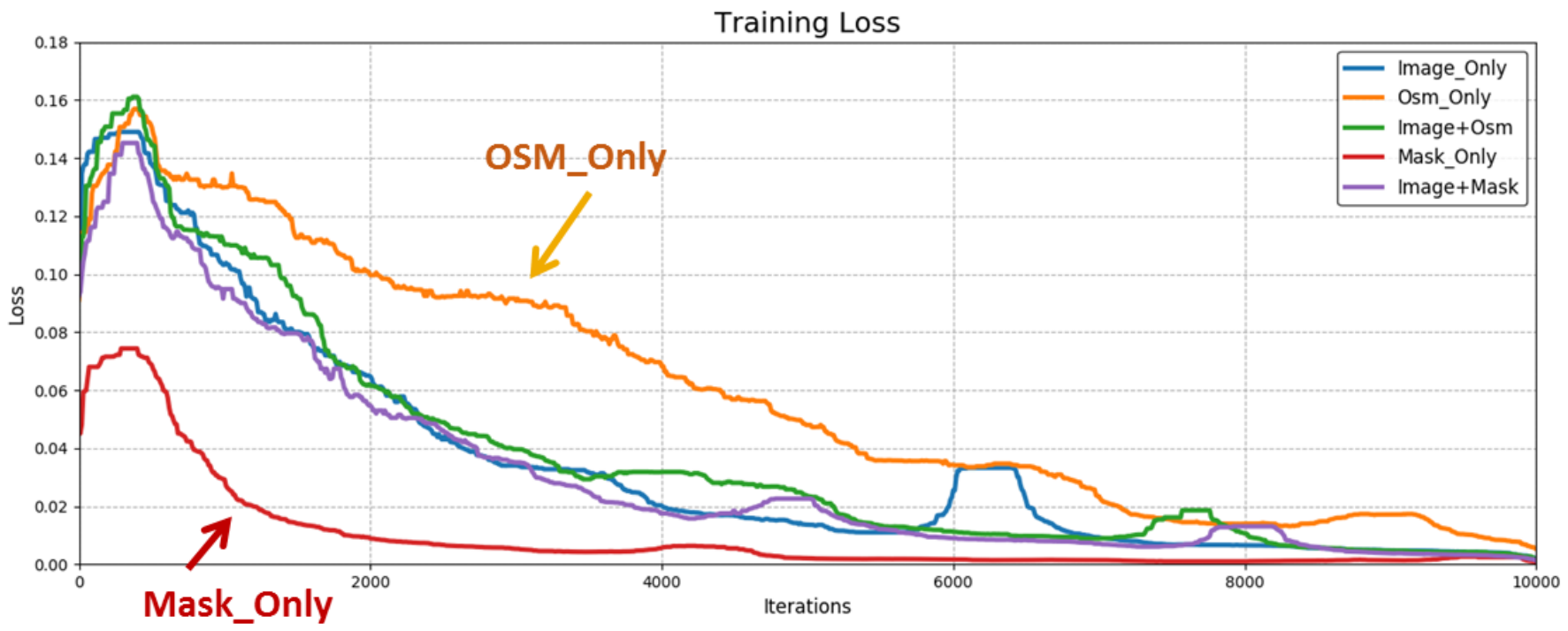

Fig. 16. Training loss decreasing history during the training iterations. Variant input types are compared in different colors.

TABLE VI
SPECIAL CASE RESULTS USING OSM IN DEEP LEARNING

**(a)**
**Training: camera**
**Annotation: human**
**Testing: OSM**

| | *Precision* | *Recall* | *F1-score* |
|---|---|---|---|
| *UM* | 0.4681 | 0.9647 | 0.6067 |
| *UMM* | **0.8025** | **0.8373** | **0.8093** |
| *UU* | 0.6191 | 0.8846 | 0.7157 |

**(b)**
**Training: camera**
**Annotation: OSM**
**Testing: camera**

| | *Precision* | *Recall* | *F1-score* |
|---|---|---|---|
| *UM* | 0.5536 | 0.8685 | 0.6477 |
| *UMM* | **0.7932** | **0.7962** | **0.7803** |
| *UU* | 0.7197 | 0.7080 | 0.6880 |

Sometimes, a large dataset is available for us to train our own model and test, however ground truth annotation may be limited and require a massive human labor. Is it possible to trust OSM and use its rendered road mask to provide the automatic annotation?

Table VI lists the experimental results for the two special cases. It is observed that UMM performs better than the others. Again, it is because the OSM rendered road mask covers all the road surface, and not a single lane, which makes this bad performance for UM. UU roads are usually in rural places with low traffic flow, which makes user contributions for OSM database limited in such areas. OSM cannot accurately estimate the unmarked lane width, and therefore performs bad for UU. Therefore, these special cases serve to probe the performance limit for OSM, but better results could be expected if a more accurate high digit map data were available.

## VIII. CONCLUSION

This study has considered the availability of using OpenStreetMap data for driving environment understanding. First, scenario attributes were shown to be parsed from OSM elements, and supplemented with vehicle dynamic signals for driving event detection. Second, OSM data was used to render virtual street views, which provided a road based prior knowledge for drivable space characterization. We obtained five types of road masks – (1) OSM super-pixel refinement approach, (2) multiple candidates approach, (3) color-based image GrabCut algorithm, (4) edge-based lane-mark approach, and (5) Lidar point cloud processing, and their combination provided the best result. An FCNN architecture, which inputs the OSM mask as an additional source, achieved a better road segmentation than the hand-crafted masks. The OSM rendered virtual street views were also expected to have the capability to provide additional test images or the automatic annotation.

As an alternative to virtual street views, OSM data can also be used to render depth layouts, which would become one more input channel in our future work to obtain road and object masks. The current road segmentation was based on single images. If a video dataset is employed in the continued work, it is expected to carry greater time series information for visual odometry object tracking. Finally, since the OSM accuracy is limited by user contributions, a more precise high digit map data is expected to provide a better description of the road prior knowledge.